\documentclass[aps,pre,twocolumn,superscriptaddress]{revtex4-1}
\usepackage{graphicx}% Include figure files
\usepackage{dcolumn}% Align table columns on decimal point
\usepackage{bm}% bold math
\usepackage{accents}
\usepackage{amssymb}
\usepackage{feynmp}
\usepackage{url}
\usepackage{setspace}
\usepackage{color}

\newcommand{\beq}{\begin{equation}}
\newcommand{\eeq}{\end{equation}}
\newcommand{\beqa}{\begin{eqnarray}}
\newcommand{\eeqa}{\end{eqnarray}}
\newcommand{\Tr}{\text{Tr}}

\begin{document}

%\preprint{}

\title{Speeding-up a quantum refrigerator via counter-diabatic driving}


\author{Ken Funo}
\email[]{ken.funo@riken.jp}
%\thanks{}
\affiliation{Theoretical Physics Laboratory, RIKEN Cluster for Pioneering Reserach, Wako-shi, Saitama 351-0198, Japan}
\author{Neill Lambert}
\affiliation{Theoretical Physics Laboratory, RIKEN Cluster for Pioneering Reserach, Wako-shi, Saitama 351-0198, Japan}
\author{Bayan Karimi}
\affiliation{QTF Centre of Excellence, Department of Applied Physics, Aalto University School of Science, Aalto, Finland}
\author{Jukka P. Pekola}
\affiliation{QTF Centre of Excellence, Department of Applied Physics, Aalto University School of Science, Aalto, Finland}
\author{Yuta Masuyama}
\affiliation{National Institutes for Quantum and Radiological Science and Technology, 1233 Watanuki, Takasaki, Gunma 370-1292, Japan}
\author{Franco Nori}
\affiliation{Theoretical Physics Laboratory, RIKEN Cluster for Pioneering Reserach, Wako-shi, Saitama 351-0198, Japan}
\affiliation{Physics Department, The University of Michigan, Ann Arbor, Michigan 48109-1040, USA}



\date{\today}

\begin{abstract}
We study the application of a  counter-diabatic driving (CD) technique to enhance the thermodynamic efficiency and power of a quantum Otto refrigerator based on a superconducting qubit coupled to two resonant circuits. Although the CD technique is originally designed to counteract non-adiabatic coherent excitations in isolated systems, we find that it also works effectively in the open system dynamics, improving the coherence-induced losses of efficiency and power. We compare the CD dynamics with its classical counterpart, and find a deviation that arises because the CD is designed to follow the energy eigenbasis of the original Hamiltonian, but the heat baths thermalize the system in a different basis. We also discuss possible experimental realizations of our model.
\end{abstract}

\maketitle

\section{Introduction}

Understanding the nonequilibrium dynamics of open quantum systems is essential for controlling small quantum devices and to improve existing quantum information processing technologies. Quantum thermodynamics offers a theoretical framework to achieve this aim, and one can, for example, study thermodynamically efficient protocols with low entropy production. Quite recently, utilizing recent technical progress in the fields of trapped ions, NMR systems, and superconducting qubits, several experiments have been performed to test important ideas in quantum thermodynamics such as the quantum fluctuation theorem~\cite{Batalhao14,An15} and Maxwell's demon~\cite{Camati16,Cottet17,Masuyama17,Naghiloo18}. They are also used as a working substance to build up quantum heat engines and refrigerators~\cite{HEexp1,HEexp2,HEexp3}. We also note that a direct measurement of the stationary heat currents has become possible~\cite{HEexp4}. 

The studies of quantum heat engines and refrigerators~\cite{Haitao1,Kosloff17} have attracted particular interest since they reveal fundamental limits on the conversion between work and heat in the quantum regime. %They are also relevant for practical applications such as cooling some particular part of the subsystem. 
%Because of their fundamental importance, many theoretical works have been performed to understand the performance of heat engines and refrigerators in the quantum regime. 
For example, several studies have found quantum supremacy in their performance~\cite{Scully03,Jaramillo16,Coherence1,Coherence2,Deffner17}. On the other hand, coherences built up during a cycle of a quantum heat engine are found to induce universal power losses in the linear response regime~\cite{Brandner17}. Similar result has also been reported in some specific models~\cite{Karimi16,Pekola18}, where coherent oscillations are found in the output power and efficiency, leading to smaller values compared to their classical counter parts.   
%For example, the ideal quantum Otto cycle consists of two quantum adiabatic strokes and two thermalization strokes. When the driving frequency becomes faster, coherent oscillations between energy eigenstates become present and the system cannot follow the ideal quantum adiabatic strokes. This leads to coherent oscillations in the output power and efficiency, having smaller values compared to their classical counter parts~\cite{Karimi16,Pekola18}. 

One may regard this as a manifestation of the trade-off between the protocol time and the efficiency of a given task in finite-time control theory. However, a recent quantum control technique, known as shortcuts to adiabaticity (STA), allows us to overcome this problem by mimicking quantum adiabatic dynamics in a finite protocol time~\cite{STAreview,STAR}. In particular, the counter-diabatic driving (CD) technique~\cite{STAreview,STAR,DR03,DR05,Berry09,Jarzynski13} realizes STA by introducing an additional control field which enforces the system to follow the quantum adiabatic trajectory of the uncontrolled system. By utilizing these techniques, the performance of superadiabatic quantum heat engines have been studied extensively~\cite{Adolfo14,Deng18,Lutz18,Berakdar16,Adolfo18}, whereas other optimization techniques have been utilized in the literature as well~\cite{Acconcia15A,Acconcia15B,Menczel19}. Note that the CD has been implemented in several experiments~\cite{expCD1,expCD2,expCD3,expCD4}.

\begin{figure}[t]
\begin{center}
\includegraphics[width=.45\textwidth]{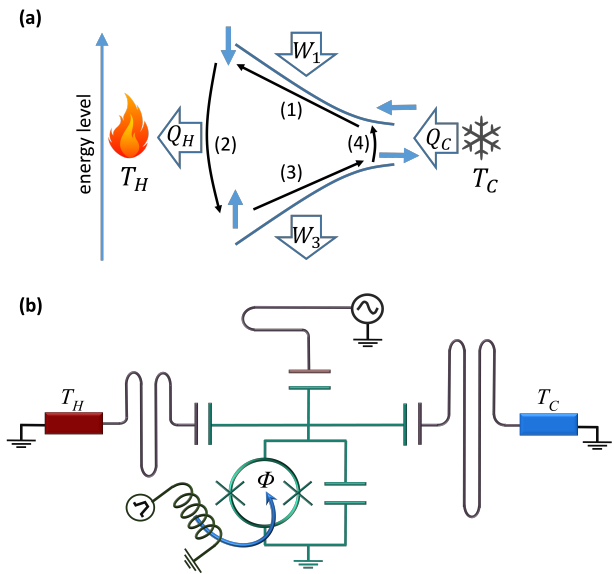}
\caption{ (a) Scheme of the quantum refrigerator studied here. (1) and (3): Adiabatic strokes by changing the energy level of the qubit. During these processes, the qubit is off-resonant with the baths and the work is supplied to or extracted from the qubit. (2) Thermalization stroke with respect to the hot bath (H). Energy is transfered from the qubit to the hot bath. (4) Thermalization stroke with respect to the cold bath (C). Energy is transfered from the cold bath to the qubit. (b) Possible experimental realizations of the quantum refrigerator using a superconducting qubit coupled to two RLC resonators and a microwave drive line. The transmon qubit Hamiltonian is given by Eq.~(\ref{Hamiltonian}), where the Josephson coupling energy (related to $q(t)$) is tuned by an externally applied magnetic flux $\Phi$. The input microwave drive realizes the counter-diabatic driving Hamiltonian Eq.~(\ref{HamiltonianCD}). The hot and cold heat baths made of RLC resonators are capacitively coupled to the qubit, and the dissipative dynamics of the system is described by Eq.~(\ref{LECD}).  
}
\label{fig_LZ}
\end{center}
\end{figure}

In this study, we take a model of a quantum Otto refrigerator based on a superconducting qubit coupled to two heat baths made of resonant circuits~\cite{Karimi16}, and apply the CD to enhance its efficiency and power. The model we consider is illustrated in Fig.~\ref{fig_LZ}, where the energy level of the qubit is varied in time and it is resonantly coupled to the hot bath (H) and the cold bath (C) at different frequencies. Note that if we can switch on and off the interactions between the system and the baths, we can separate the adiabatic strokes and the thermalization strokes of the Otto engine (see also Fig.~\ref{fig_LZ}). Then, we can ideally apply the CD to speed up the adiabatic strokes~\cite{Adolfo14,Deng18}. On the other hand, we are interested in a situation where the coupling to the baths cannot be externally controlled and the adiabatic and thermalization strokes are not completely separated. From a practical point of view, this setup is relevant for realistic experiments where the system undergoes a continuous periodic cycle with some external drives under the influence of environments. From a fundamental point of view, this setup allows one to better understand how CD could be effective in open system dynamics, which has not been explored intensively~\cite{Sun16,Sarandy17,Villazon19}. 

This paper is organized as follows. In Sec.~\ref{sec:model}, we present the model studied in this paper describing a quantum Otto refrigerator. We also introduce the CD technique and the definition of the work flux and the heat flux for our model. The main result of our paper is presented in Sec.~\ref{sec:Main}. We first discuss some analytical expression for the dynamics of the system and show that the CD also works effectively for the open quantum system of this example. We then discuss how CD improves the heat transfer and the thermodynamic efficiency of the refrigerator in the fast driving regimes. In Sec.~\ref{sec:Exp}, we discuss possible experimental realizations of the quantum refrigerator studied in this paper. In Sec.~\ref{sec:Con}, we conclude this paper.  

\begin{figure}[t]
\begin{center}
\includegraphics[width=.45\textwidth]{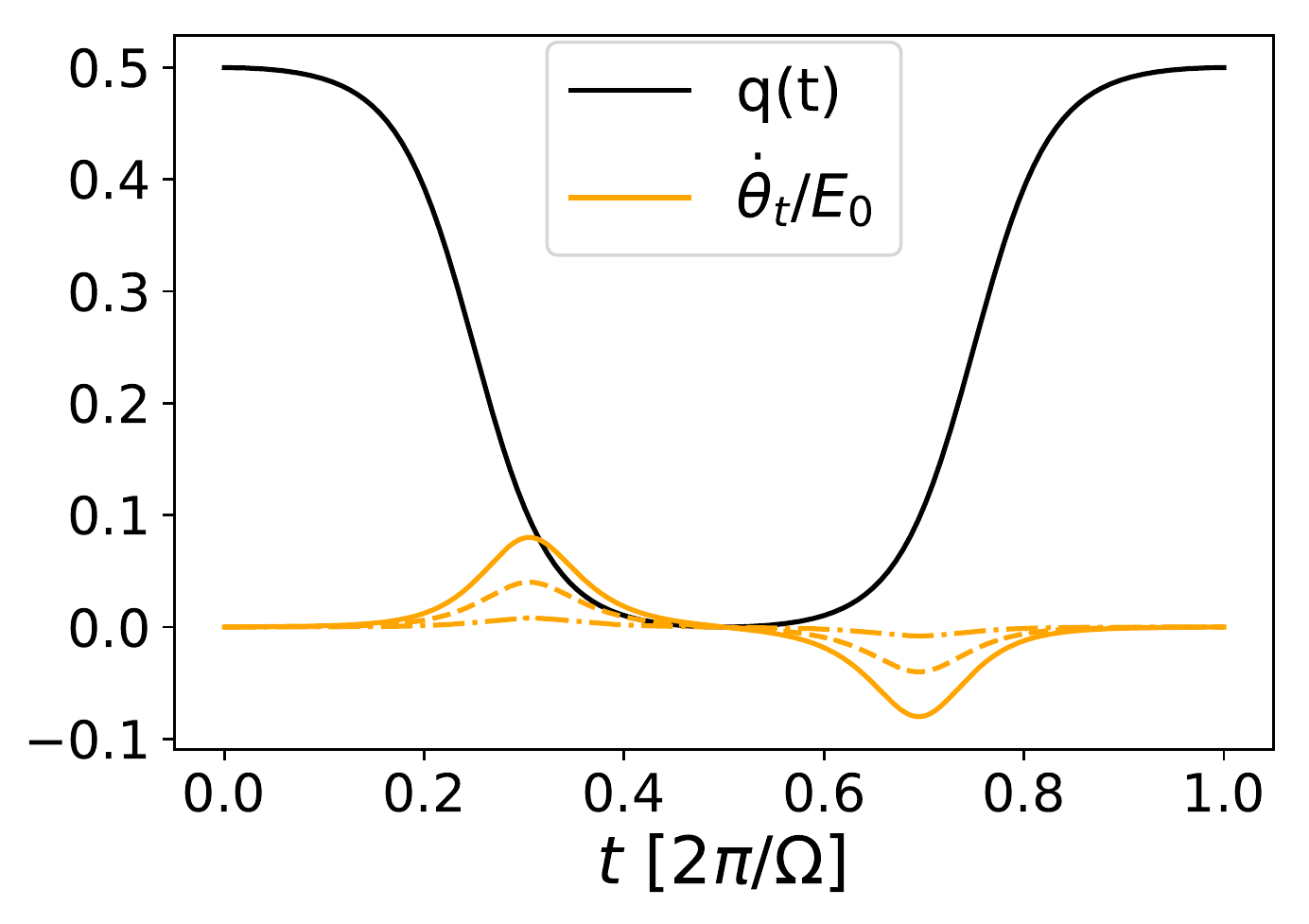}
\caption{ Functional form of the drivings $q(t)$~(\ref{q}) [black curve] and $\dot{\theta}_{t}$~(\ref{dtheta}) [orange curves] for one cycle. Here, $\dot{\theta}_{t}$ is plotted for $\Omega=0.1$, $\Omega=0.05$, and $\Omega=0.01$ (from top to bottom). Note that the amplitude of $\dot{\theta}_{t}$ is proportional to $\Omega$. We choose the parameters $a=2$, $E_{0}=1$ and $\Delta = 0.12$.      
}
\label{fig_driving}
\end{center}
\end{figure}

\section{\label{sec:model}The model}

The Hamiltonian of the qubit is given by the Landau-Zener-type model
\beq
H_{0}(t)=-E_{0}(\Delta \sigma_{x}+q(t)\sigma_{z}) , \label{Hamiltonian}
\eeq
where $E_{0}$ is the overall energy of the qubit, $\Delta$ characterizes the minimum gap, $q(t)$ describes the external driving, and $\sigma_{x}$ is the Pauli-X-matrix, etc. Here, we choose $q(t)$ as a periodic function varying from $q=0$ to $q=1/2$. We choose the truncated trapezoidal form
\beq
q(t)=\frac{1}{4}\left(1+\frac{\tanh(a\cos\Omega t)}{\tanh(a)}\right), \label{q}
\eeq
which in earlier works was shown to give the best thermodynamic efficiency among several different functional forms~\cite{Karimi16}. Here, $\Omega$ is the driving frequency and $a$ is a parameter adjusting the waveform of the periodic drive (see Fig.~\ref{fig_driving}). The energy difference between the excited state and the ground state is given by 
\beq
\Delta \epsilon(t)= 2E_{0}\sqrt{\Delta^{2}+q^{2}(t)}.
\eeq
The instantaneous eigenenergies of $H_{0}$ are given by $\epsilon_{\rm{e/g}}(t)=\pm \Delta \epsilon/2$, and the corresponding energy eigenstates are given by
\beqa
|\epsilon_{\rm{e}}(t)\rangle&=&\cos\theta_{t}\left|\uparrow\right\rangle +\sin\theta_{t}\left|\downarrow \right\rangle, \nonumber \\
|\epsilon_{\rm{g}}(t)\rangle&=&\sin\theta_{t}\left|\uparrow\right\rangle -\cos\theta_{t}\left|\downarrow \right\rangle,
\eeqa
where $\theta_{t}=(1/2)\cot^{-1}(q/\Delta)$.
%\beq
%\theta_{t}=\frac{1}{2}\cot^{-1}\left(\frac{q(t)}{\Delta}\right).
%\eeq

Now we consider the dissipative dynamics of the system coupled to the hot and cold baths.  The coupling between the system and the bath $i=\{\rm{C},\rm{H}\}$ are assumed to take the forms $H_{\rm{int}}^{i}=\sigma_{y}\otimes B^{i}$, where $B^{i}$ is the operator of the bath $i$ (including the coupling constant). Note that we discuss the case of a $\sigma_{y}$ (transversal) coupling between the system and the bath $i$ (see also Fig.~\ref{fig_LZ} (b)), although a $\sigma_{z}$ (longitudinal) coupling does not significantly change the qualitative behavior of the results presented in this paper.  After taking the standard weak-coupling, Born-Markov, and rotating-wave approximations, the reduced dynamics of the system is given by the time-dependent Lindblad master equation~\cite{Breuer,Albash12,Yuge17}
\beq
\partial_{t}\rho = -i[H_{0},\rho]+D^{\rm{C}}[\rho]+D^{\rm{H}}[\rho], \label{LEoriginal}
\eeq
where we set $\hbar=1$ for simplicity. Here, the dissipator describing the effect of the bath $i$ is given by
\beqa
D^{i}[\rho]&=&S^{i}(\Delta \epsilon)\left[ L\rho L^{\dagger}-\frac{1}{2}\{L^{\dagger}L,\rho\}\right] \nonumber \\
& &+ S^{i}(-\Delta \epsilon)\left[ L^{\dagger}\rho L-\frac{1}{2}\{ LL^{\dagger},\rho\}\right], \label{Dissioriginal}
\eeqa
where $\{A,B\}=AB+BA$ denotes the anti-commutation relation and 
\beq
L=|\epsilon_{\rm{g}}\rangle \langle \epsilon_{\rm{g}}|\sigma_{y}|\epsilon_{\rm{e}}\rangle \langle \epsilon_{\rm{e}}| = i|\epsilon_{\rm{g}}\rangle \langle \epsilon_{\rm{e}}|
\eeq
is the time-dependent Lindblad operator describing a jump from the excited state to the ground state. %Note that we discuss the case of a $\sigma_{y}$ (transversal) coupling between the system and the bath $i$ (see also Fig.~\ref{fig_LZ} (b)), although a $\sigma_{z}$ (longitudinal) coupling does not significantly change the qualitative behavior of the results presented in this paper. %[we also use the relation $S_{-\Delta \epsilon}=L^{\dagger}$ in Eq.~(\ref{Dissioriginal})]. 
 Here, $S^{i}(\omega)$ is the noise power spectrum of the environment and it is related to the one-sided Fourier transform of the bath correlation function $g^{i}(\tau)=\langle B^{i}(\tau)B^{i}(0) \rangle$ as $\int^{\infty}_{0}d\tau e^{i\omega\tau}g^{i}(\tau)=\frac{1}{2}S^{i}(\omega)+i\lambda^{i}(\omega)$. Note that we ignore the Lamb shift term $\lambda^{i}(\omega)$ in Eq.~(\ref{LEoriginal}) for simplicity. 

In this paper, we consider the following form of the noise power spectrum: 
\beq
S^{i}(\Delta\epsilon)=\frac{g_{i}}{2}\frac{1}{1+Q_{i}^{2}(\Delta\epsilon/\omega_{i}-\omega_{i}/\Delta\epsilon)^{2}}\frac{\Delta\epsilon}{1-\exp(-\beta_{i}\Delta\epsilon)}, \label{noisepower}
\eeq
 since it is relevant to the possible experimetnal realizations of the refrigerator~\cite{Karimi16} (see also Fig.~\ref{fig_LZ} (b)). Here, $\omega_{i}=1/\sqrt{L_{i}C_{i}}$, $Q_{i}=R_{i}^{-1}\sqrt{L_{i}/C_{i}}$, $L_{i}$, $C_{i}$, $R_{i}$, $\beta_{i}$ and $g_{i}$ are the bare resonance frequency, quality factor, inductance, capacitance, resistance, inverse temperature, and coupling strength of the circuit $i=\{\rm{C},\rm{H}\}$, respectively. We choose $\omega_{\rm{C}}=2E_{0}\Delta$ and $\omega_{\rm{H}}=2E_{0}\sqrt{\Delta^{2}+1/4}$, such that the circuit C (H) is resonantly coupled to the qubit when $q=0$ ($q=1/2$), where $Q_{i}$ adjusts the width of the resonance.  
%We denote the energy difference of the qubit at $q=1/2$ and $q=0$ as
%\beqa
%\hbar\omega_{\rm{C}}&=&2E_{0}\sqrt{1/4+\Delta^{2}}, \nonumber \\
%\hbar\omega_{\rm{H}}&=&2E_{0}\Delta,
%\eeqa
%respectively. 

%We also have
%\beq
%S_{-\Delta \epsilon}=S_{\Delta\epsilon}^{\dagger}
%\eeq

%We also have
%\beqa
%\dot{\theta}_{t}&=&-\frac{\dot{\mathcal{Q}}}{2}\frac{\Delta}{\Delta^{2}+q^{2}}, \\
%\theta_{t}&=&\frac{1}{2}\cot^{-1}\left(\frac{q}{\Delta}\right).
%\eeqa

\subsection{Counter-diabatic driving}
In this subsection, we briefly introduce the idea of CD and then apply it to our model. 

By following Ref.~\cite{Jarzynski13}, we introduce the control field $H_{1}(t)$ to escort the state along the same label $n$ of the energy eigenstate of $H_{0}(t)$ as 
\beq
|\epsilon_{n}(t)\rangle \rightarrow \left(1-i\delta t H_{1}(t)\right)|\epsilon_{n}(t)\rangle = e^{i\delta t A_{n}(t)}|\epsilon_{n}(t+\delta t)\rangle, \label{adtransport}
\eeq
and $A_{n}(t)=i\langle \epsilon_{n}(t)|\partial_{t}\epsilon_{n}(t)\rangle$ is the Berry connection. This means $H_{1}$ transports the state along the quantum adiabatic trajectory $|\epsilon_{n}(t)\rangle$ for the original Hamiltonian $H_{0}$. Here, the control field can be obtained from Eq.~(\ref{adtransport}) and its explicit form is given by
\beq
H_{1}(t)= i\sum_{n} (1-|\epsilon_{n}\rangle\langle \epsilon_{n}|)| \partial_{t}\epsilon_{n}\rangle\langle \epsilon_{n}|, \label{CD}
\eeq
which is called the Counter-Diabatic (CD) field~\cite{STAR,DR03,DR05,Berry09}. As one can expect from Eq.~(\ref{adtransport}), the unitary time-evolution $U_{\rm{cd}}=\Tr[\exp(-i\int^{t}_{0}ds H_{\rm{cd}}(s))]$, via the Hamiltonian $H_{\rm{cd}}=H_{0}+H_{1}$, mimics the quantum adiabatic time-evolution of $H_{0}$ in a finite time $t$ as
\beq
U_{\rm{cd}}=\sum_{n}e^{i\int^{t}_{0} ds (A_{n}(s)-\epsilon_{n}(s))}|\epsilon_{n}(t)\rangle\langle\epsilon_{n}(0)| . \label{UCD}
\eeq

Now we apply the CD technique to our model~(\ref{Hamiltonian}). Since $|\partial_{t}\epsilon_{\rm{g}}\rangle=\dot{\theta}_{t}|\epsilon_{\rm{e}}\rangle$ and $|\partial_{t}\epsilon_{\rm{e}}\rangle=-\dot{\theta}_{t}|\epsilon_{\rm{g}}\rangle$, the CD field takes a simple form
\beq
H_{1}=\dot{\theta}_{t}\sigma_{y}, \label{HamiltonianCD}
\eeq
with
\beq
\dot{\theta}_{t}=-\frac{\dot{q}}{2}\frac{\Delta}{\Delta^{2}+q^{2}}. \label{dtheta}
\eeq
Note that $\dot{\theta}_{t}$ is proportional to $\Omega$ (see also Fig.~\ref{fig_driving}). 
%\beqa
%H_{1}&=&i\hbar( \left|\partial_{t}\epsilon_{\rm{e}}\rangle\langle \epsilon_{\rm{e}}\right|+\left|\partial_{t}\epsilon_{\rm{g}}\rangle\langle \epsilon_{\rm{g}}\right|) \nonumber \\
%&=&i\hbar\dot{\theta}_{t}(-\left|\epsilon_{\rm{g}}\rangle\langle \epsilon_{\rm{e}}\right|+\left|\epsilon_{\rm{e}}\rangle\langle \epsilon_{\rm{g}}\right|) \nonumber \\
%&=&\hbar\dot{\theta}\sigma_{y}\nonumber \\
%&=&-\frac{\hbar\dot{\mathcal{Q}}}{2}\frac{\Delta}{\Delta^{2}+q^{2}}\sigma_{y}
%\eeqa
%and thus
%\beq
%H_{\rm{cd}}(t)=-E_{0}(\Delta\sigma_{x}+q(t)\sigma_{z})-\frac{\hbar \dot{\mathcal{Q}}(t)}{2}\frac{\Delta }{\Delta^{2}+q^{2}(t)}\sigma_{y}.
%\eeq
The energy difference between the excited state and the ground state of $H_{\rm{cd}}$ is given by 
\beq
\Delta \epsilon_{\rm cd} = 2\sqrt{E_{0}^{2}(\Delta^{2}+q^{2})+\dot{\theta}_{t}^{2}}.
\eeq 
%The eigenenergies are given by $\epsilon^{\rm cd}_{\rm{e/g}}(t)=\pm\Delta \epsilon_{\rm cd}/2$ and
%\beqa
%E_{\pm}(t)&=&\pm\sqrt{E_{0}^{2}(\Delta^{2}+q^{2})+\hbar^{2}\dot{\theta}^{2}} \nonumber \\
%&=&\pm \sqrt{E_{0}^{2}(\Delta^{2}+q^{2})+\frac{\hbar^{2}\dot{\mathcal{Q}}^{2}}{4}\frac{\Delta^{2}}{(\Delta^{2}+q^{2})^{2}} }.
%\eeqa
%we denote the corresponding eigenstates by $|\epsilon^{\rm cd}_{\rm{e/g}}\rangle$. 
%given by
%\beqa
%|\epsilon^{\rm cd}_{\rm{e}}\rangle&=&\cos\psi \left|\uparrow\right\rangle+e^{i\Phi}\sin\psi\left|\downarrow\right\rangle \nonumber\\
%|\epsilon^{\rm cd}_{\rm{g}}\rangle &=&-\sin\psi\left|\uparrow\right\rangle+e^{i\Phi}\cos\psi\left|\downarrow\right\rangle
%\eeqa
%with the coefficients
%\beqa
%\psi&=&\frac{1}{2}\cos^{-1}\left( \frac{2E_{0}q}{\Delta \epsilon_{\rm cd}}\right) \nonumber \\
%\Phi&=&\tan^{-1}\left( \frac{E_{0}\Delta}{\sqrt{E_{0}^{2}\Delta^{2}+\hbar^{2}\dot{\theta}^{2}}} \right).
%\eeqa

Next, we consider the time-dependent master equation~\cite{Breuer,Albash12,Yuge17} including the CD field, given by
\beq
\partial_{t}\rho_{\rm{cd}}=-i[H_{0}+H_{1},\rho_{\rm{cd}}]+\mathcal{D}^{\rm{C}}[\rho_{\rm{cd}}]+\mathcal{D}^{\rm{H}}[\rho_{\rm{cd}}], \label{LECD}
\eeq
where
%\beqa
%\mathcal{D}^{i}[\rho_{\rm{cd}}]&=&S^{i}(\Delta \epsilon_{\rm cd})\left[ L_{\rm cd}\rho_{\rm{cd}} L^{\dagger}_{\rm cd}-\frac{1}{2}\{L^{\dagger}_{\rm cd}L_{\rm cd},\rho_{\rm{cd}}\}\right] \nonumber \\
%& +&S^{i}(-\Delta \epsilon_{\rm cd})\left[ L^{\dagger}_{\rm cd}\rho_{\rm{cd}} L_{\rm cd}- \frac{1}{2}\{ L_{\rm cd}L^{\dagger}_{\rm cd},\rho_{\rm{cd}}\}\right] \nonumber 
%\eeqa
%is the dissipator and the time-dependent Lindblad operator describing a jump from the excited state to the ground state is given by
the dissipator $\mathcal{D}^{i}$ is given by Eq.~(\ref{Dissioriginal}) but replacing $\Delta\epsilon$ and $L$ by $\Delta \epsilon_{\rm cd}$ and 
\beq
L_{\rm cd}:=\left| \epsilon^{\rm cd}_{\rm{g}}\rangle\langle \epsilon^{\rm cd}_{\rm{e}}\right| \left\langle \epsilon^{\rm cd}_{\rm{g}}|\sigma_{y}|\epsilon^{\rm cd}_{\rm{e}}\right\rangle %=\frac{E_{0}\Delta +i\dot{\theta}}{(\Delta \epsilon_{\rm cd}/2)}\left| \epsilon^{\rm cd}_{\rm{g}}\rangle\langle \epsilon^{\rm cd}_{\rm{e}}\right|, \label{LindbladOPCD}
\eeq
where $|\epsilon_{\rm g}^{\rm cd}\rangle$ and $|\epsilon_{\rm e}^{\rm cd}\rangle$ are the ground and excited state of $H_{\rm cd}$, respectively. 
%Here, note that dissipation is acting in the direction of the eigenbasis of $H_{\rm{cd}}$, opposed to Eqs.~(\ref{LEoriginal}). 

 % and (\ref{LEpert}). 
%We later discuss that this discrepancy leads to the reduction of the efficiency of the refrigerator via CD compared with its classical counterpart.
%We now introduce the ground and excited occupation probabilities as $P^{\rm cd}_{\rm{g}}(t)=\langle \epsilon^{\rm cd}_{\rm{g}}|\rho_{\rm{cd}}|\epsilon^{\rm cd}_{\rm{g}}\rangle$ and $P^{\rm cd}_{\rm{e}}(t)=\langle \epsilon^{\rm cd}_{\rm{e}}|\rho_{\rm{cd}}|\epsilon^{\rm cd}_{\rm{e}}\rangle$, respectively, and also introduce the off-diagonal element as $\sigma_{\rm{ge}}=\langle \epsilon^{\rm cd}_{\rm{g}}|\rho_{\rm{cd}}|\epsilon^{\rm cd}_{\rm{e}}\rangle$. %Then, the Lindblad master equation~(\ref{LECD}) can be rewritten as
%\beqa
%\partial_{t}P^{\rm cd}_{\rm{g}}&=& \Gamma_{\downarrow}P^{\rm cd}_{\rm{e}}-\Gamma_{\uparrow}P^{\rm cd}_{\rm{g}} , \\
%\partial_{t}\sigma_{\rm{ge}}&=& ,
%\eeqa

%\subsection{Classical model}
%If we consider the time-evolution of the diagonal components $p_{\pm}:=\langle E_{\pm}|\rho|E_{\pm}\rangle$ of the density matrix, we have the following Pauli master equation
%\beq
%\left\langle \epsilon^{\rm cd}_{\rm{g}}|D[\rho]|\epsilon^{\rm cd}_{\rm{g}}\right\rangle = 2\Gamma(\Delta \epsilon_{\rm cd})\frac{E^{2}_{0}\Delta^{2}+\hbar^{2}\dot{\theta}^{2}}{(\Delta \epsilon_{\rm cd})^{2}}\rho_{ee}-2\Gamma(-\Delta \epsilon_{\rm cd})\frac{E^{2}_{0}\Delta^{2}+\hbar^{2}\dot{\theta}^{2}}{(\Delta \epsilon_{\rm cd})^{2}}\rho_{gg}.
%\eeq

\subsection{\label{sec:heatflux}Heat fluxes to the cold and hot baths}
In this section, we introduce the expression of the heat fluxes from the cold and hot baths for the original~(\ref{LEoriginal}) and CD~(\ref{LECD}) dynamics. 

For the original dynamics without CD, the time-derivative of the internal energy of the system is given by $\dot{E}=\Tr[ (\partial_{t}H_{0})\rho]+\Tr[H_{0}(\partial_{t}\rho)]$. From the first law of thermodynamics $\dot{E}=\dot{W}-\dot{\mathcal{Q}}$, we identify $\dot{W}=\Tr[ (\partial_{t}H_{0})\rho(t)]$ as the work flux, since this term characterizes the energy difference of the system induced by the external driving of the Hamiltonian. Similarly, we identify the term $\dot{\mathcal{Q}}=-\Tr[ H_{0}(\partial_{t}\rho)]=\Tr[H_{0}(D^{\rm{C}}[\rho]+D^{\rm{H}}[\rho])]$ as the heat flux and further decompose it into two parts $\dot{\mathcal{Q}}=\dot{\mathcal{Q}}^{\rm{C}}+\dot{\mathcal{Q}}^{\rm{H}}$, where
\beq
\dot{\mathcal{Q}}^{i}=\Delta \epsilon\left[ \Gamma_{\downarrow,i}(t)P_{\rm{e}}(t) - \Gamma_{\uparrow,i}(t)P_{\rm{g}}(t) \right]   \label{defQflux}
\eeq
is the heat flux coming from the bath $i$%=\{\rm{C},\rm{H}\}$
. Here,  $P_{\rm{g}}(t)=\langle \epsilon_{\rm{g}}|\rho|\epsilon_{\rm{g}}\rangle$ is the ground state occupation probability and $P_{\rm{e}}(t)=\langle \epsilon_{\rm{e}}|\rho|\epsilon_{\rm{e}}\rangle$ is that for the excited state, and the transition rates are given by
\beq
\Gamma_{\downarrow,i}(t)=S^{i}(\Delta \epsilon),\ \Gamma_{\uparrow,i}(t)=S^{i}(-\Delta \epsilon). \label{Trates1}
\eeq
%and we use the notation $\gamma_{\uparrow}=\gamma_{\uparrow,1}+\gamma_{\uparrow,2}$ and  $\gamma_{\downarrow}=\gamma_{\downarrow,1}+\gamma_{\downarrow,2}$. 
It is clear from the expression~(\ref{defQflux}) that any change in the energy of the system related to a jump between energy eigenstates induced by the bath is interpreted as the heat. We note that the Gibbs equilibrium state $\rho^{\rm G}_{\beta_{i}}$ with inverse temperature $\beta_{i}$ satisfies $D^{i}[\rho^{\rm G}_{\beta_{i}}]=0$ and $-i[H_{0},\rho^{\rm G}_{\beta_{i}}]=0$. Therefore, the above definitions of the work and the heat for the Lindblad master equation dynamics are consistent with the second law of thermodynamics~\cite{Alicki79}.

For the CD dynamics, we can define the heat flux in a manner similar to that for the original dynamics by replacing $H_{0}$ with $H_{\rm{cd}}$. After some similar arguments, the heat flux from the system to the bath $i$ %=\{\rm{C},\rm{H}\}$
 is found to be
%The time-derivative of the internal energy of the system is given by $\dot{E}=\Tr[ (\partial_{t}H_{\rm{cd}})\rho]+\Tr[H_{\rm{cd}}(\partial_{t}\rho)]$. From the first law of thermodynamics $\dot{E}=\dot{W}-\dot{\mathcal{Q}}$, we identify $\dot{W}=\Tr[ (\partial_{t}H_{\rm{cd}})\rho(t)]$ as the work flux since this term characterizes the energy difference of the system induced by the external driving of the Hamiltonian. Similarly, we identify the term $\dot{\mathcal{Q}}=-\Tr[ H_{\rm{cd}}(\partial_{t}\rho_{\rm{cd}})]=\Tr[H_{\rm{cd}}\mathcal{D}_{\rm{cd}}[\rho_{\rm{cd}}]]$ as the heat flux and further decompose it into two parts $\dot{\mathcal{Q}}=\dot{\mathcal{Q}}_{\rm{C}}+\dot{\mathcal{Q}}_{\rm{H}}$, where
\beq
\dot{\mathcal{Q}}^{i}_{\rm{cd}}=\Delta \epsilon_{\rm cd}\left[ \Gamma_{\downarrow,i}^{\rm cd}(t)P^{\rm cd}_{\rm{e}}(t) - \Gamma_{\uparrow,i}^{\rm cd}(t)P^{\rm cd}_{\rm{g}}(t) \right]  , \label{defQfluxCD}
\eeq
where $P^{\rm cd}_{\rm{g}}(t)=\langle \epsilon^{\rm cd}_{\rm{g}}|\rho_{\rm{cd}}|\epsilon^{\rm cd}_{\rm{g}}\rangle$ is the ground state occupation probability and a similar definition applies to $P^{\rm cd}_{\rm{e}}(t)$, %$P^{\rm cd}_{\rm{e}}(t)=\langle \epsilon^{\rm cd}_{\rm{e}}|\rho_{\rm{cd}}|\epsilon^{\rm cd}_{\rm{e}}\rangle$ is that for the excited state, 
while the transition rates are given by
\beqa
\Gamma_{\downarrow,i}^{\rm cd}(t)&=& \left( \frac{\Delta\epsilon}{\Delta \epsilon_{\rm cd}}\right)^{2}S^{i}(\Delta \epsilon_{\rm cd}), \nonumber \\
 \Gamma_{\uparrow,i}^{\rm cd}(t)&=&\left( \frac{\Delta\epsilon}{\Delta \epsilon_{\rm cd}}\right)^{2}S^{i}(-\Delta \epsilon_{\rm cd}), \label{TratesCD}
\eeqa
%for $i=\{\rm{C},\rm{H}\}$. 
Note that when the driving $\Omega$ is slow, $\dot{\theta}$ becomes negligible and Eqs.~(\ref{TratesCD}) and~(\ref{Trates1}) become identical.

\section{\label{sec:Main}Main results: Heat fluxes and the thermodynamic efficiency of the Otto cycle}
We now numerically solve the Lindblad master equation and calculate the heat flux as well as the efficiency of the Otto refrigerator, which constitute our main results.  

\subsection{Dynamics of the Otto cycle}

%\begin{figure}[ht]
%\begin{center}
%\includegraphics[width=.45\textwidth]{images/up_P005.pdf}
%\caption{ $|\uparrow\rangle$ state population.   
%}
%\label{fig1}
%\end{center}
%\end{figure}

\begin{figure}[t]
\begin{center}
\includegraphics[width=.45\textwidth]{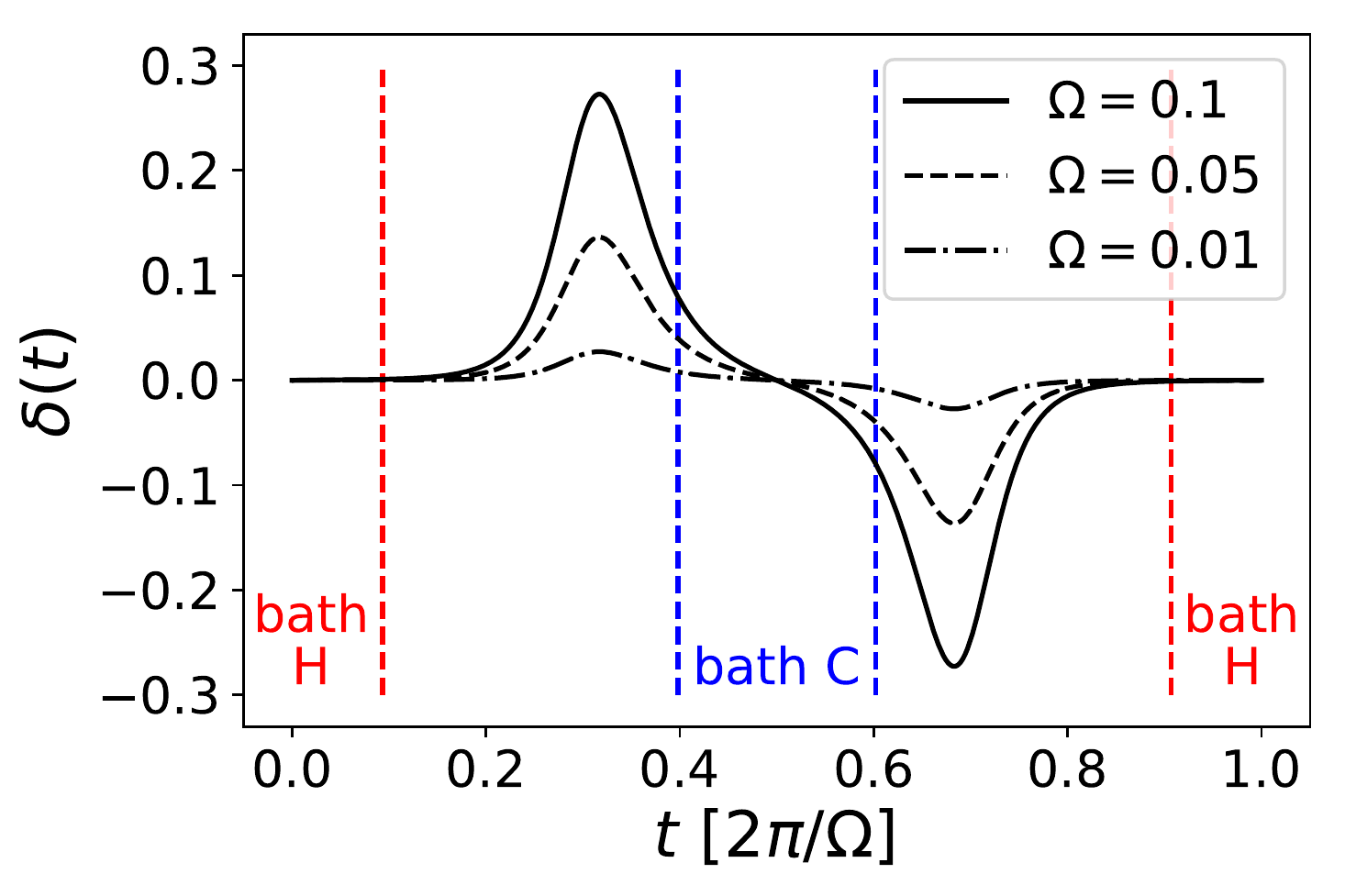}
\caption{ Functional form of the relative energy scale of the CD field with respect to the original Hamiltonian $\delta(t)=\dot{\theta}/\Delta\epsilon$~(\ref{delta}) for different driving frequency $\Omega$ [black curves]. The vertical red (blue) dashed lines indicate the time region in which the interaction between the hot (cold) bath and the qubit is dominant. Note that the CD works well if $\delta(t)$ is sufficiently small during the time region in which the system interacts with the heat baths (see Eq.~(\ref{CDdelta})). From this figure, we find that the CD-assisted control is affected by the cold bath, and the performance of the refrigerator is degraded. The parameters used here are $a=2$, $E_{0}=1$, $\Delta=0.12$, $\beta_{\rm{C}}^{-1}=0.15$, $\beta_{\rm{H}}^{-1}=0.3$, $g_{\rm{C}}=g_{\rm{H}}=1$, and $Q_{\rm{C}}=Q_{\rm{H}}=30$.  
}
\label{fig:delta}
\end{center}
\end{figure}

\begin{figure}[t]
\begin{center}
\includegraphics[width=.45\textwidth]{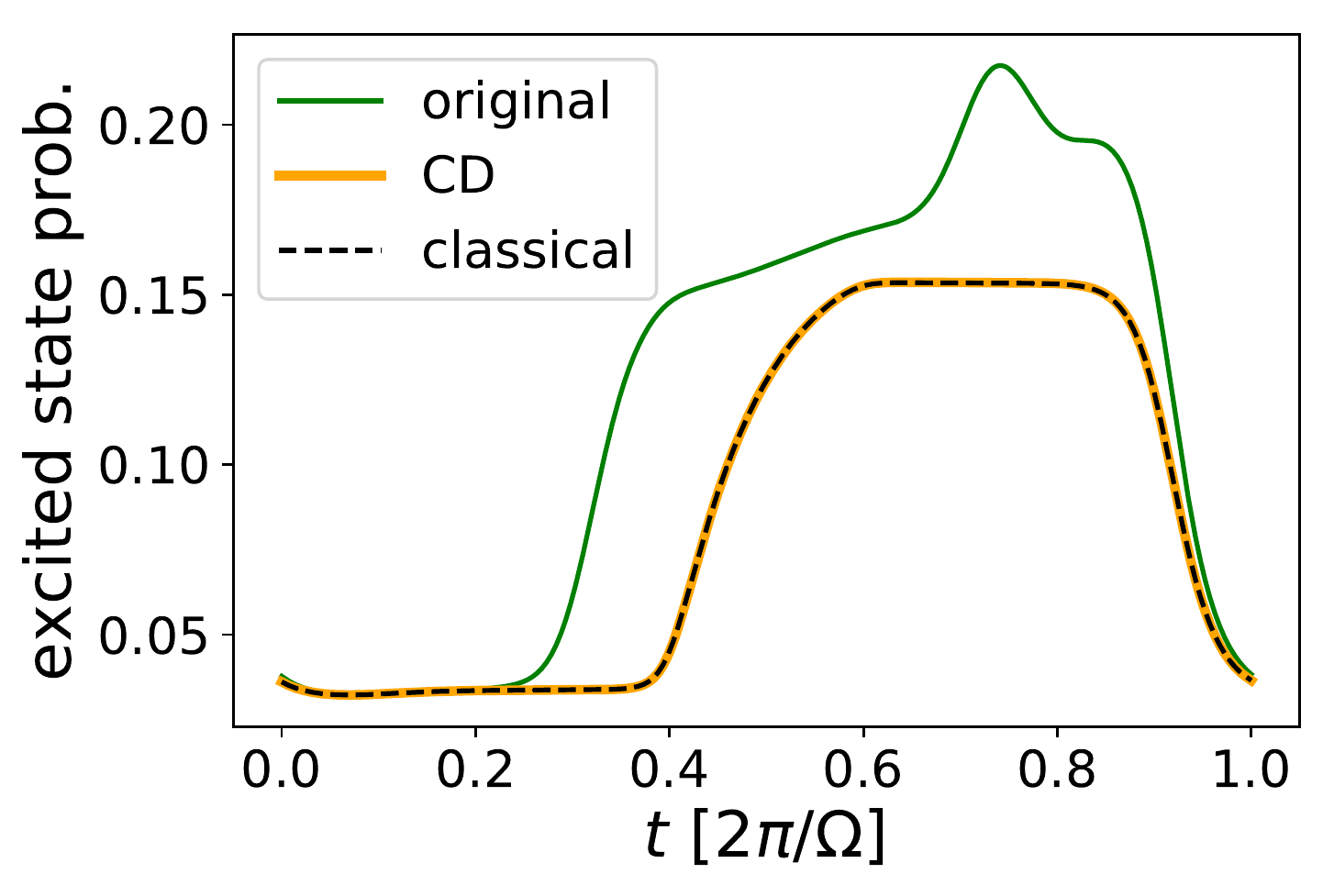}
\caption{ Excited state probability using the basis $|\epsilon_{\rm e}\rangle$ for the CD %($\langle \epsilon_{\rm e}|\rho_{\rm cd}|\epsilon_{\rm e}\rangle$)
($P^{\rm cd}_{|\epsilon_{\rm e}\rangle}$) [orange solid curve], original ($P_{\rm{e}}$) [green solid curve] and classical [black dashed curve] dynamics. %Here, we choose the driving frequency as $\Omega=0.05$, and the system does not have enough time to completely thermalize. The excited state probability is $P=1.68$ ($P=0.03$) when the system is completely thermalized with C (H). 
Note that the excited state probability for the CD dynamics is almost identical to that of the classical dynamics, showing the effectiveness of the CD technique even in open quantum systems. The parameters used here are $\Omega=0.1$, $a=2$, $E_{0}=1$, $\Delta=0.12$, $\beta_{\rm{C}}^{-1}=0.15$, $\beta_{\rm{H}}^{-1}=0.3$, $g_{\rm{C}}=g_{\rm{H}}=1$, and $Q_{\rm{C}}=Q_{\rm{H}}=30$. 
}
\label{fig2}
\end{center}
\end{figure}

\begin{figure}[t]
\begin{center}
\includegraphics[width=.45\textwidth]{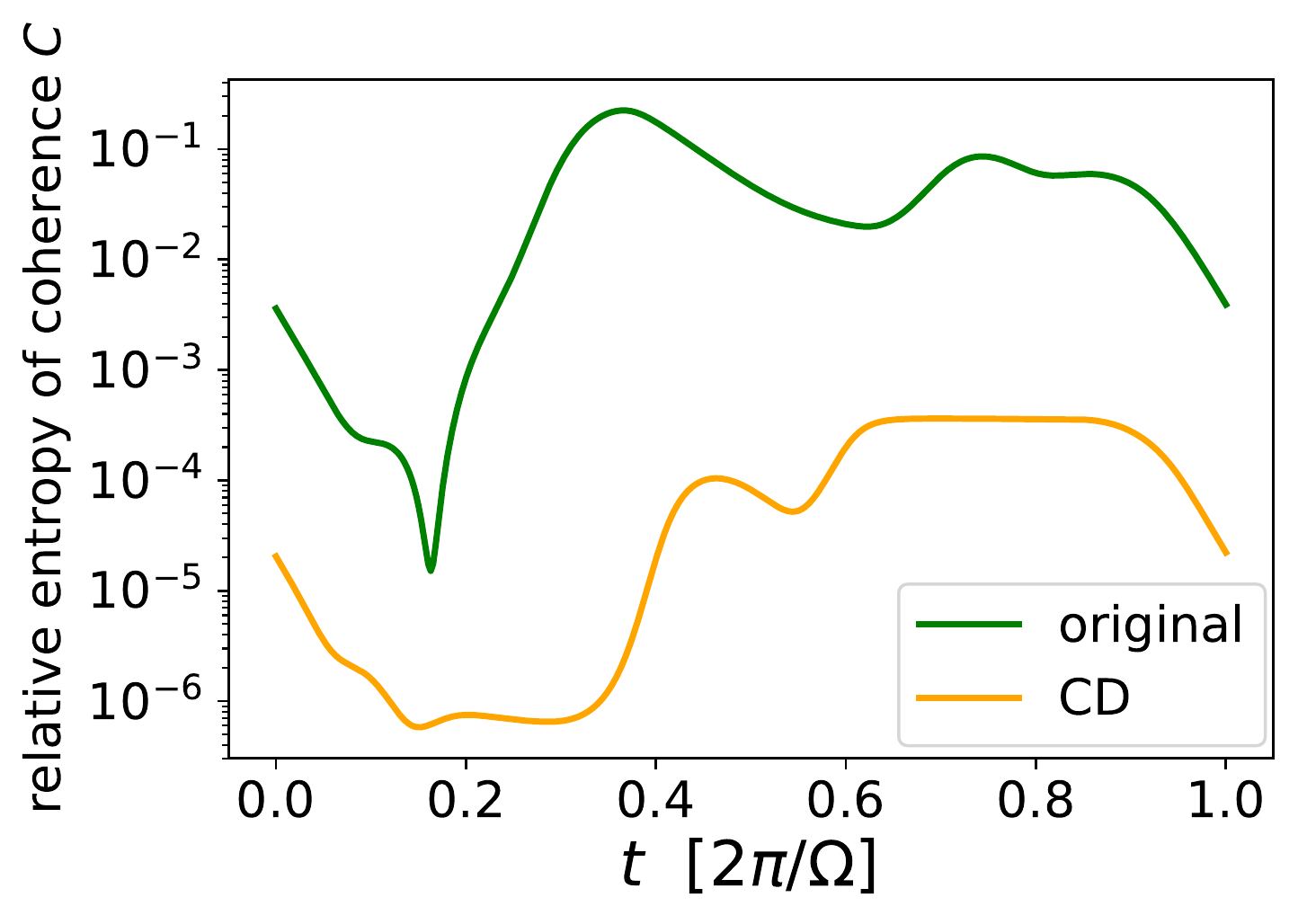}
\caption{ Relative entropy of coherence $C$ for the CD dynamics $C(\rho_{\rm{cd}})$ [orange curve] and the original dynamics $C(\rho)$ [green  line]. Note that the coherence between different energy eigenstates $|\epsilon_{n}\rangle$ is suppressed by at least one order of magnitude via the CD. This suppression improves the coherence induced losses of power and efficiency. 
%(a) $C(\rho_{\rm{cd}})$ for different driving frequencies $\Omega$. (b) $C(\rho)$ for different driving frequencies $\Omega$. 
%Here, $C$ quantifies the coherence between different energy eigenstates $|\epsilon_{n}\rangle$ of the original Hamiltonian $H_{0}$. 
The parameters are $\Omega=0.1$, $a=2$, $E_{0}=1$, $\Delta=0.12$, $\beta_{\rm{C}}^{-1}=0.15$, $\beta_{\rm{H}}^{-1}=0.3$, $g_{\rm{C}}=g_{\rm{H}}=1$, and $Q_{\rm{C}}=Q_{\rm{H}}=30$. 
}
\label{figC}
\end{center}
\end{figure}

%Figure~\ref{fig1} plots the state of the qubit projected to the basis $\left| \uparrow\right\rangle$ for different driving frequencies $\Omega$. At low frequency regime, the dynamics of the system with and without CD (red and black solid lines, respectively) are found to be almost identical. This is because $\dot{\theta}$ is small and the original dynamics without CD already follows the quantum adiabatic dynamics quite well when the system is off-resonant with respect to the heat baths. We note that the system does not completely relax to the thermal state when it interacts with the heat bath and we find some deviation from the instantaneous stationary state shown by the dotted blue line. When $\Omega$ becomes larger as in Fig.~\ref{fig1} (c), we find coherent oscillations for the original dynamics (black solid line) as reported in Ref.~\cite{Karimi16}. On the other hand, CD cancels these oscillations and let the system follow the stationary state when it is off-resonant, as shown by the red solid line. 

%We first note that the design of the protocol allows $\dot{\theta}\simeq 0$ when $q\simeq 0$ or $q\simeq 1/2$ (see Fig.~\ref{fig_driving}). This ensures that the system with CD resonantly couples to the bath at the points $q=0$ and $q=1/2$, similar to the original protocol without CD. It also means that $\Delta\epsilon\simeq \Delta \epsilon_{\rm cd}$, and thus the dissipation mainly acts in the basis $|\epsilon_{n}\rangle$, allowing the CD to work well even for open quantum systems. 

We first note that the design of the protocol lets the CD field become small ($\dot{\theta}\simeq 0$) when the qubit is interacting with the hot or cold baths ($q\simeq 0$ or $q\simeq 1/2$). This ensures that the CD is less affected by the baths and is able to cancel nonadiabatic excitations during the cycle. As a result, the coherence between different energy eigenstates of the original Hamiltonian is supressed, and the coherence induced power and efficiency losses~\cite{Brandner17} can be avoided.  

To support this idea, let us denote the matrix elements of $\rho_{\rm cd}$ using the basis $|\epsilon_{n}\rangle$ as 
\beqa
P^{\rm cd}_{|\epsilon_{n}\rangle}&=&\langle\epsilon_{n}|\rho_{\rm{cd}}|\epsilon_{n}\rangle , \\
\delta\rho^{\rm cd}_{\rm ge}&=&\langle \epsilon_{\rm g}|\rho_{\rm cd}|\epsilon_{\rm e}\rangle.
\eeqa
The Lindblad master equation~(\ref{LECD}) can be rewritten as a Pauli master equation-like form
\beqa
\partial_{t}P^{\rm cd}_{|\epsilon_{\rm g}\rangle}&=&\sum_{i}\left(\Gamma^{\rm cd}_{\downarrow,i}P^{\rm cd}_{|\epsilon_{\rm e}\rangle}-\Gamma^{\rm cd}_{\uparrow,i}P^{\rm cd}_{|\epsilon_{\rm g}\rangle}\right)+O(\delta^{2})+O(\delta\rho_{\rm ge}^{\rm cd},\delta), \nonumber \\
\partial_{t}\delta\rho^{\rm cd}_{\rm ge}&=& -\frac{1}{2}\sum_{i}\left(\Gamma^{\rm cd}_{\downarrow,i}+\Gamma^{\rm cd}_{\uparrow,i}\right)\delta\rho^{\rm cd}_{\rm ge}+O(\delta), \label{CDdelta}
\eeqa
where 
\beq
\delta(t)=\frac{\dot{\theta}_{t}}{\Delta \epsilon(t)} \label{delta}
\eeq
quantifies the relative energy scale of the CD field with respect to the original Hamiltonian (see Fig.~\ref{fig:delta}). We therefore find that if $\delta\rho_{\rm ge}(0)=0$ and the driving frequency $\Omega$ is not too large such that $\delta$ is small, the CD dynamics is essentially described by the classical master equation (i.e., the first line of Eq.~(\ref{CDdelta}) by neglecting $O(\delta^{2})$ and $O(\delta\rho^{\rm cd}_{\rm ge},\delta)$ terms). However, in general, we cannot completely cancel the coherent excitations because there is a mismatch between the basis $|\epsilon_{n}\rangle$ in which the CD is designed to follow and the basis $|\epsilon^{\rm cd}_{n}\rangle$ in which the dissipation acts on.  Note that this mismatch is quantified by $|\langle \epsilon^{\rm cd}_{\rm{g}}|\epsilon_{\rm{g}}\rangle|^{2}=1/2+\Delta\epsilon/(2\Delta \epsilon_{\rm cd})=1-\delta^{2}+O(\delta^{4})$. 

%In Fig.~\ref{fig2}, we plot the excited state population probability $P^{\rm cd}_{|\epsilon_{\rm{e}}\rangle}$.  %using the basis $|\epsilon_{\rm{e}}\rangle$ for the original dynamics ($P^{\rm cd}_{\rm{e}}$) and the CD ($P^{\rm cd}_{|\epsilon_{\rm{e}}\rangle}$).  

In Fig.~\ref{fig2}, we plot $P^{\rm cd}_{|\epsilon_{\rm{e}}\rangle}$, which shows an excellent agreement with that of the classical model. On the other hand, we find coherent oscillations for the original dynamics $P_{\rm e}$. We further consider the effectiveness of CD by analyzing the coherence of the system between different energy eigenstates $|\epsilon_{n}\rangle$. We adopt the relative entropy of coherence $C(\sigma)=S(\sigma^{\rm{d}})-S(\sigma)$ for a density matrix $\sigma$, which is found to be a proper measure of coherence~\cite{Baumgratz14}. Here, $S(\sigma)=-\Tr[\sigma\ln\sigma]$ is the von Neumann entropy and $\sigma^{\rm{d}}=\sum_{n}|\epsilon_{n}\rangle\langle\epsilon_{n}|\sigma|\epsilon_{n}\rangle\langle\epsilon_{n}|$ is the diagonal part of $\sigma$. Note that when $C(\sigma)=0$, $\sigma$ has no coherence between eigenstates $|\epsilon_{n}\rangle$. In Fig.~\ref{figC}, we plot the relative entropy of coherence for the CD [$C(\rho_{\rm{cd}})$] and original [$C(\rho)$] dynamics, and find that $C(\rho_{\rm{cd}})$ is at least one order of magnitude smaller than $C(\rho)$. %showing that CD is able to counteract non-adiabatic coherent excitations even in open quantum systems. 

\begin{figure}[t]
\begin{center}
\includegraphics[width=.45\textwidth]{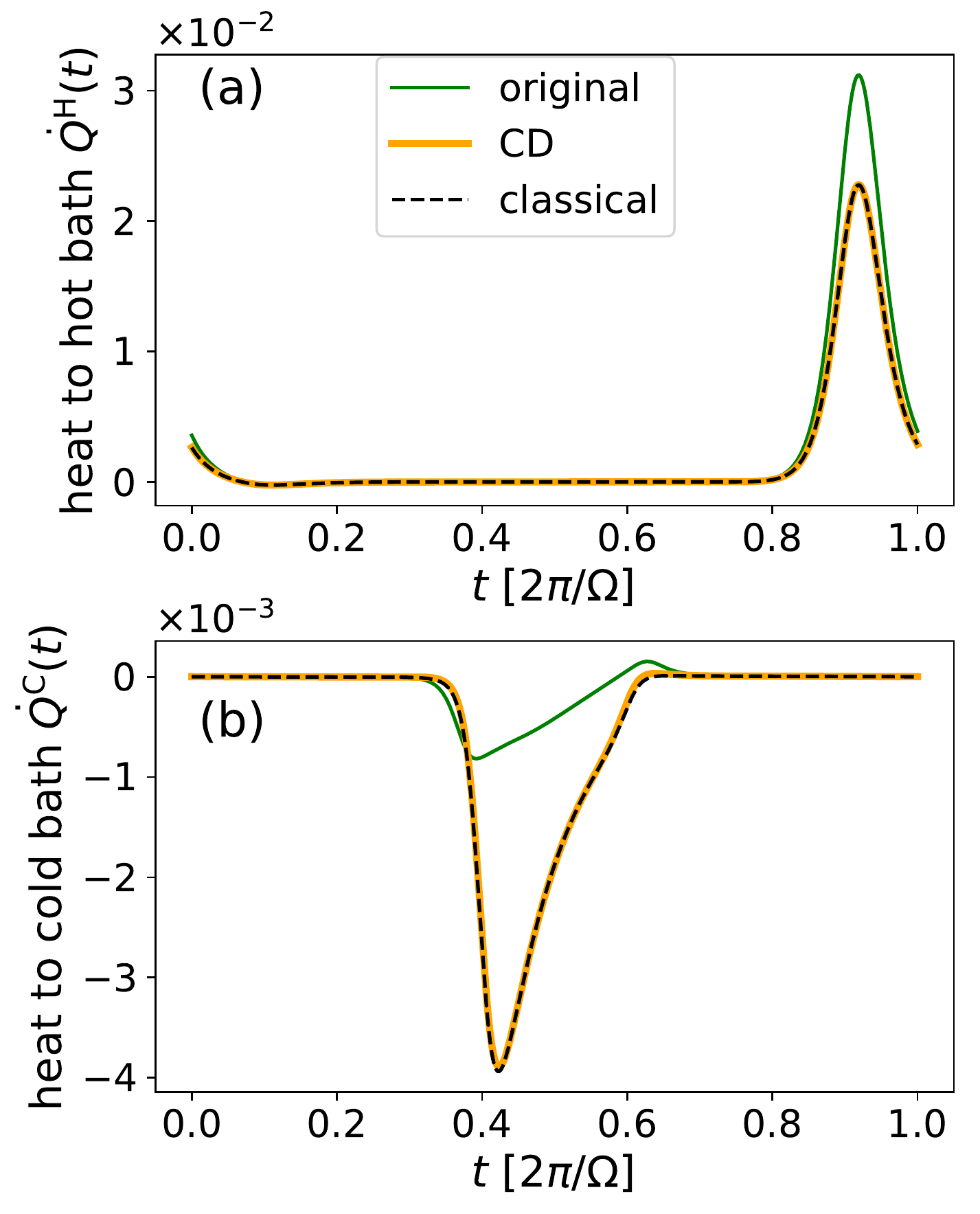}
\caption{ Heat fluxes $\dot{\mathcal{Q}}$ as functions of time $t$ for one cycle. (a) Heat flux to the hot bath for the original dynamics $\dot{\mathcal{Q}}^{\rm{H}}$ [green solid curve], CD dynamics $\dot{\mathcal{Q}}^{\rm{H}}_{\rm cd}$ [orange solid curve], and classical dynamics [black dashed curve].  (b) Heat flux to the cold bath ($\dot{\mathcal{Q}}^{\rm C}$). Note that the negative value of $\dot{\mathcal{Q}}^{\rm C}$ means that the energy is transfered from the cold bath to the system. We also note that the system exchanges heat with the hot and cold baths around the resonance points $q(t)=1/2$ and $q(t)=0$, respectively. These plots show that the original protocol is not working effectively, compared with the classical model, to transport heat from the cold bath to the hot bath. However, the CD technique largely improves the efficiency of transporting heat since the heat fluxes between the CD and classical dynamics are almost identical. The parameters are $\Omega=0.1$, $a=2$, $E_{0}=1$, $\Delta=0.12$, $\beta_{\rm{C}}^{-1}=0.15$, $\beta_{\rm{H}}^{-1}=0.3$, $g_{\rm{C}}=g_{\rm{H}}=1$, and $Q_{\rm{C}}=Q_{\rm{H}}=30$.  
}
\label{fig3}
\end{center}
\end{figure}

%\begin{figure}[ht]
%\begin{center}
%\includegraphics[width=.45\textwidth]{images/heat_flux005.pdf}
%\caption{ Here, $\Omega=0.05$.   
%}
%\label{fig3a}
%\end{center}
%\end{figure}

\subsection{Heat flux between the system and the two heat baths}

Next, we study the heat flux. Here, the sign convention of the heat is chosen such that when it is positive, heat flows from the system to the bath. In Fig.~\ref{fig3} (a), we plot the heat flux to the hot bath, where the interaction is dominant around $q=1/2$. Here, the heat flux $\dot{\mathcal{Q}}_{\rm{cd}}^{\rm{H}}$ via CD has an excellent agreement with its classical counterpart $\dot{\mathcal{Q}}_{\rm{cl}}^{\rm{H}}$, calculated from the classical master equation. This agreement can be understood from Fig.~\ref{fig:delta} that $\delta\simeq 0$ when the system is interacting with the hot bath ($q\simeq 1/2$). %Here, the classical heat flux is given by
%\beq
%\dot{\mathcal{Q}}_{\rm{cl}}^{i}=\Delta \epsilon_{\rm cd} (\Gamma_{\downarrow,i}^{\rm cd}P^{\rm{cl}}_{\rm{e}}-\Gamma_{\uparrow,i}^{\rm cd}P^{\rm{cl}}_{\rm{g}}). \label{defQfluxCL}
%\eeq 
In Fig.~\ref{fig3} (b), we plot the heat flux to the cold bath, where the interaction is dominant around $q=0$. Here, the heat flux $\dot{\mathcal{Q}}_{\rm{cd}}^{\rm{C}}$ agrees well with its classical counterpart $\dot{\mathcal{Q}}_{\rm{cl}}^{\rm{C}}$, although we find a slight deviation because $\delta$ is finite when the system is interacting with the cold bath ($q\simeq 0$). See also Fig.~\ref{fig:delta}.  %This deviation arises from the mismatch between the basis $|\epsilon_{n}\rangle$ in which the CD is designed to follow and the basis $|\epsilon^{\rm cd}_{n}\rangle$ in which the dissipation acts on. 
%Note that the overlap $|\langle \epsilon^{\rm cd}_{\rm{g}}|\epsilon_{\rm{g}}\rangle|^{2}=1/2+\Delta\epsilon/(2\Delta \epsilon_{\rm cd})$ is smaller when the system is interacting with the bath C ($q\simeq 0$) than with the bath H ($q\simeq 1/2$).

%By looking at the value of $\dot{\theta}$ in Fig.~\ref{fig_driving}, we find that the deviation between those bases is much larger when the system is interacting with the cold bath ($q\simeq 0$) than with the hot bath ($q\simeq 1/2$). This explains why we see a slight difference between CD and classical results in Fig.~\ref{fig3} (b) but not in (a).

The heat fluxes for the original dynamics $\dot{\mathcal{Q}}^{\rm{H}}$ and $\dot{\mathcal{Q}}^{\rm{C}}$ [green solid curve] take different values compared with the classical model because of the coherent oscillations shown in Fig.~\ref{fig2} and Fig.~\ref{figC}. When $\Omega$ is too large, the heat from the cold bath may change sign, i.e. the cold bath is heated up.  

\begin{figure}[t]
\begin{center}
\includegraphics[width=.45\textwidth]{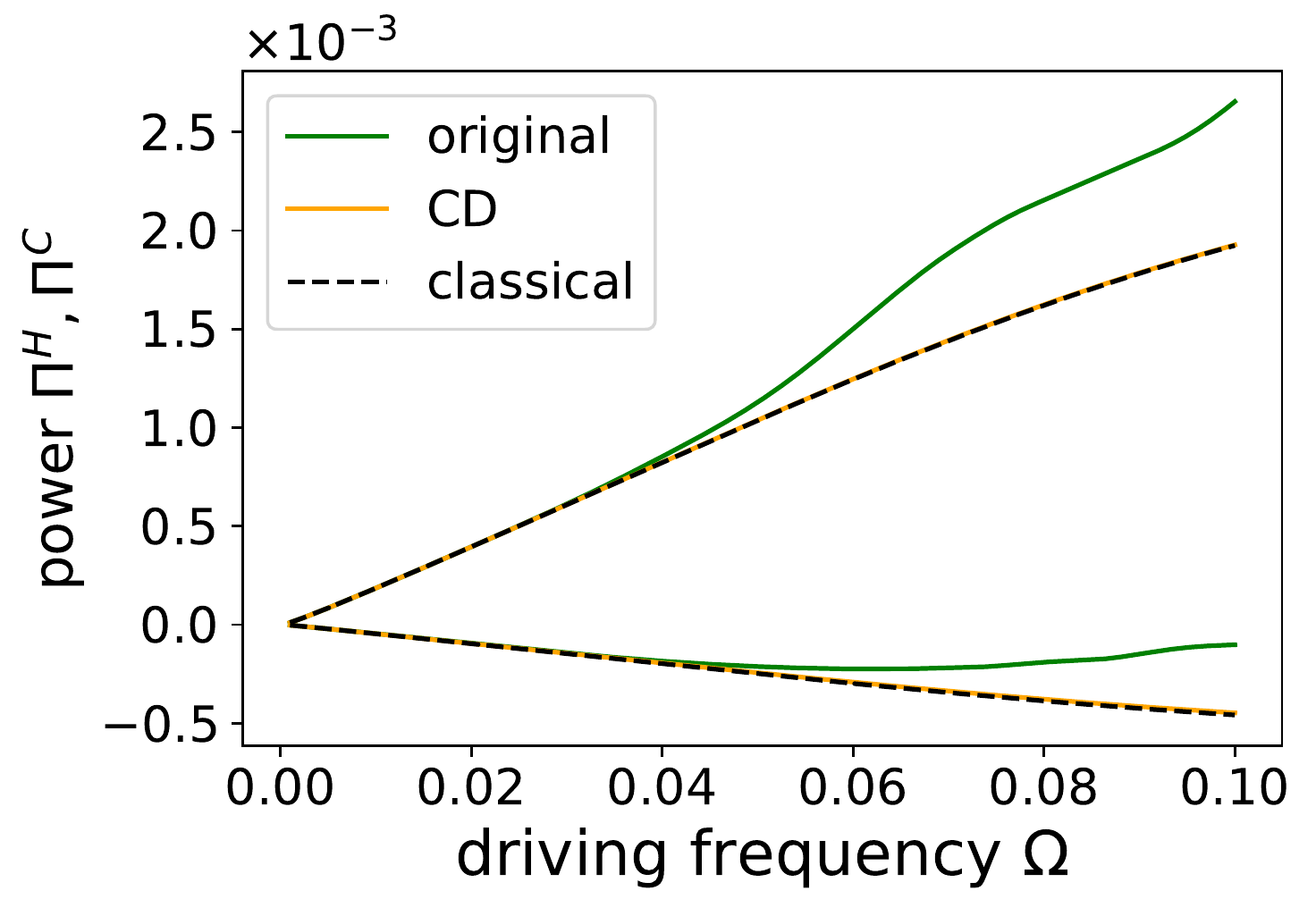}
\caption{ Cooling power $\Pi^{\rm C}$ (descending curves) of the cold bath and the heating power $\Pi^{\rm H}$ (acending curves) of the hot bath as functions of the driving frequency $\Omega$.  These plots show that for the original protocol, in the large $\Omega$ regime the cooling of the cold bath is degraded while the hot bath is more heated up. On the other hand, the CD technique largely improves the cooling power as well as suppressing the heating power. We note that the cooling and heating powers of the CD dynamics are almost identical (but have tiny differences) to those of the classical model.  The parameters are $a=2$, $E_{0}=1$, $\Delta=0.12$, $\beta_{\rm{C}}^{-1}=0.15$, $\beta_{\rm{H}}^{-1}=0.3$, $g_{\rm{C}}=g_{\rm{H}}=1$, and $Q_{\rm{C}}=Q_{\rm{H}}=30$. 
}
\label{fig_power}
\end{center}
\end{figure}

\begin{figure}[t]
\begin{center}
\includegraphics[width=.45\textwidth]{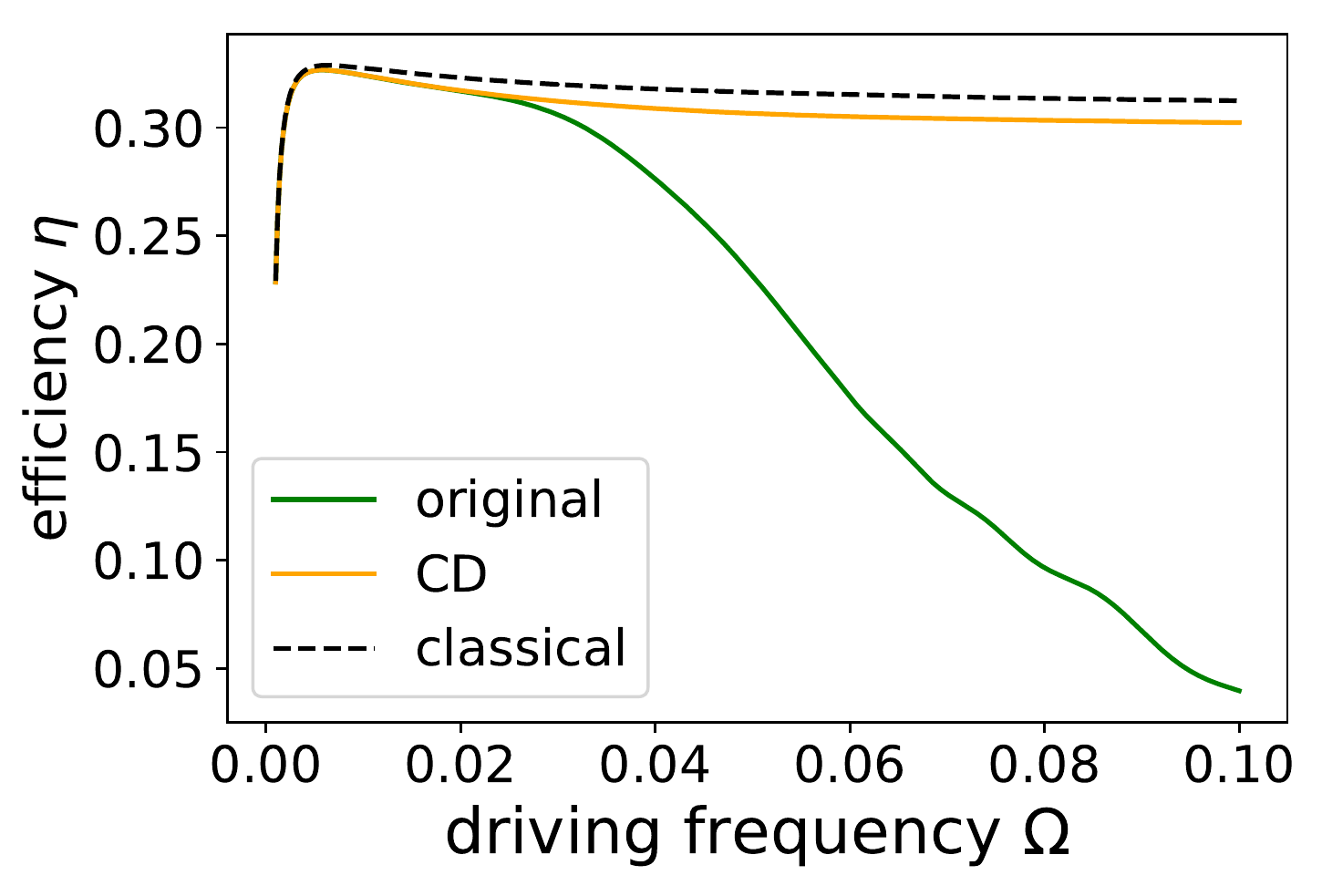}
\caption{ Thermodynamic efficiency of the refrigerator as a function of the driving frequency $\Omega$. Note that in the large $\Omega$ regime, the efficiency of the original dynamics is significantly decreased because of the coherent induced losses. On the other hand, the CD technique largely improves the efficiency. Here, the Carnot efficiency is $\eta=(\beta_{\rm C}/\beta_{\rm H}-1)^{-1}=1$ and the conventional Otto efficiency is $\eta=(\omega_{\rm H}/\omega_{\rm C}-1)^{-1}=0.304$. Note again that in our model, the quantum adiabatic strokes and the thermalization strokes are not completely seprarated because of the finite width of the noise power spectrum~(\ref{noisepower}). As a result, in the quasistatic limit ($\Omega\rightarrow 0$), our model does not recover the conventional Otto cycle and the efficiency drops down since the ``quantum adiabatic stroke'' is strongly affected by the bath. The parameters are $a=2$, $E_{0}=1$, $\Delta=0.12$, $\beta_{\rm{C}}^{-1}=0.15$, $\beta_{\rm{H}}^{-1}=0.3$, $g_{\rm{C}}=g_{\rm{H}}=1$, and $Q_{\rm{C}}=Q_{\rm{H}}=30$. 
}
\label{fig4}
\end{center}
\end{figure}

\subsection{Thermodynamic efficiency of the refrigerator}
Finally, we compare the power and the thermodynamic efficiency (coefficient of performance) of the refrigerator. The efficiency of the original dynamics is given by $\eta=-\mathcal{Q}^{\rm{C}}/W=-\mathcal{Q}^{\rm{C}}/(\mathcal{Q}^{\rm{H}}+\mathcal{Q}^{\rm{C}})$, where we use the first law of thermodynamics $W=\mathcal{Q}^{\rm{H}}+\mathcal{Q}^{\rm{C}}$ for a stationary cycle and obtain the second equality, and $\mathcal{Q}^{i}=\int dt \dot{\mathcal{Q}}^{i}(t)$ and $W=\int dt \dot{W}(t)$ are the heat and work for one stationary cycle, respectively. The efficiencies for the CD dynamics and the classical dynamics are defined in a similar manner. 
Note that we include the effect of the CD in a standard manner by defining the efficiency based on the total Hamiltonian including the CD field. We also note that there are several proposals for the energy costs of  STA~\cite{Santos15,Campbell17,Funo17}, including a modified definition of the efficiency~\cite{Lutz18}.   The cooling power of the cold bath is defined as $\Pi^{\rm C}=\mathcal{Q}^{\rm C}/(2\pi/\Omega)$, and a similar definition applies to the heating power $\Pi^{\rm H}$ as well. 

We plot the power in Fig.~\ref{fig_power} and the efficiency in Fig.~\ref{fig4} as a function of the driving frequency $\Omega$ for the original dynamics, CD dynamics, and classical dynamics. Because of the coherent oscillations seen in Fig.~\ref{fig2} and Fig.~\ref{figC} for the original dynamics, the population of the ground and excited states may be reversed and $\mathcal{Q}^{\rm{C}}$ varies from negative to positive values depending on $\Omega$. This affects the cooling power and the efficiency as it falls down rapidly in the large $\Omega$ regime. For the CD dynamics, we can largely improve them in the large $\Omega$ regime. For the cooling and heating powers, we find that the differences between the CD dynamics and the classical dynamics are tiny. However, these differences become apparent in the efficiency, as we find a slight decrease of the efficiency for the CD dynamics compared with that for the classical dynamics. Since $\delta$ scales linearly in $\Omega$, the discrepancy of the efficiency between the CD and classical dynamics becomes larger as we speed up the thermodynamic cycle.

 %We also calculate the efficiency for the classical dynamics but using $\Delta\epsilon$ and $\Gamma_{\uparrow/\downarrow,i}$ instead of $\Delta\epsilon^{\rm cd}$ and $\Gamma_{\uparrow/\downarrow,i}^{\rm cd}$. 
% replacing $\Gamma_{\uparrow,\downarrow}$ with $\gamma_{\uparrow,\downarrow}=\gamma^{\rm{C}}_{\uparrow,\downarrow}+\gamma^{\rm{H}}_{\uparrow,\downarrow}$. 
%This model is a classical counterpart of the original dynamics~(\ref{LEoriginal}), although we could not find a significant difference of $\eta$ between the two classical models.

\section{\label{sec:Exp}Experimental Feasibility}
Finally, we discuss possible experimental realizations of the refrigerator cycles proposed in this paper. The qubit Hamiltonian $H_{0}$~(\ref{Hamiltonian}) can be realized by a transmon qubit, where the external magnetic flux $\Phi(t)$ is applied to the SQUID-loop and the Josephson coupling energy $E_{J}[\Phi]$ is tunable (See Fig.~\ref{fig_LZ}). In this case, $q(t)$ is given by $q=(\Phi-\Phi_{0}/2)/\Phi_{0}$, where $\Phi_{0}=h/2e$ is the superconducting flux quantum. The energy gap at $q=0$ is characterized by $\Delta\sim E_{C}/E_{J}[\Phi_{0}/2]$ and the overall energy is $E_{0}\sim E_{J}[\Phi_{0}/2]$, where $E_{C}$ refers to the Cooper pair charging energy. 

The CD field $H_{1}$~(\ref{HamiltonianCD}) can be realized by the standard $x,y$-axis single-qubit rotation, where a microwave drive line is capacitively coupled to the qubit (see Fig.~\ref{fig_LZ}). The interaction Hamiltonian reads $\Omega_{\rm d}V_{\rm d}(t)\sigma_{y}$, where $\Omega_{\rm d}$ is the qubit-microwave coupling frequency and $V_{\rm d}(t)$ is the time-dependent voltage which is applied to the qubit through the microwave drive line~\cite{transmonreview}. By choosing $\dot{\theta}_{t}=\Omega_{\rm d}V_{\rm d}(t)$, the CD field $H_{1}$ (\ref{HamiltonianCD}) can be implemented.

The $\sigma_{y}$ coupling of the qubit to the hot and cold baths can be realized by capacitively coupling the qubit to two resonators (See Fig.~\ref{fig_LZ}). We note that a transmon qubit has been capacitively coupled to two RLC resonators (without modulating the qubit frequency) and the stationary heat currents have been measured experimentally~\cite{HEexp4}.

%We note that a standard Lindblad master equation used to describe the dissipation of a qubit which is driven by a classical input field is given by Eq.~(\ref{LEoriginal}) and treating $H_{1}$ as a perturbation: 
%\beq
%\partial_{t}\rho = -i[H_{0}+H_{1},\rho]+D^{\rm{C}}[\rho]+D^{\rm{H}}[\rho]+({\rm resonator}), \label{LEpert}
%\eeq
%where (resonator) denotes terms related to the interaction Hamiltonian between the qubit and the resonator as well as the Hamiltonian and dissipator of the resonator field. We then find that $P^{\rm cd}_{|\epsilon_{n}\rangle}$ exactly follows the classical master equation apart from the (resonator) terms, and this quantifies the difference of the thermodynamic efficiency between the classical model and the CD model with this particular realization~(\ref{LEpert}). However, we note that the applicability of the time-dependent Lindblad master equation is quite subtle and we should either compare the model with the experiment or to use an exact model such as the spin-boson model.

We also note that $H_{0}+H_{1}$ can be realized in various information processing systems by driving the qubit with classical fields in the $\sigma_{x}$, $\sigma_{y}$ and $\sigma_{z}$ directions in order to realize the $E_{0}\Delta \sigma_{x}$, $\dot{\theta}_{t}\sigma_{y}$ and $E_{0}q(t)\sigma_{z}$ terms. Note that this technique is standard in many quantum information experiments such as superconducting qubits~\cite{SCQ,transmonreview}, NMR systems~\cite{NMR}, and NV-center spins~\cite{NV}, where one can rotate the qubit in any direction of the Bloch sphere. It has also been utilized to generate a time-dependent Hamiltonian and its control CD field for a superconducting Xmon qubit~\cite{CDXmon}.

\section{\label{sec:Con}Concluding remarks}
In conclusion, we have studied the performance of a quantum Otto-type refrigerator assisted by the counter-diabatic driving (CD) technique. We find that the CD can effectively counteract non-adiabatic coherent excitations even in open quantum systems, allowing a large improvement of the thermodynamic efficiency of the refrigerator. A comparison with a classical model is also studied, and we show the deviation of the CD dynamics from the classical master equation in terms of a parameter $\delta(t)$~(\ref{delta}) which quantifies the relative energy scale between the CD field and the original Hamiltonian. This deviation arises from the mismatch between the basis in which the dissipation acts on and that in which the CD is designed to follow, and decreases the performance of the CD.  We have also discussed experimental feasibility of the proposed quantum refrigerator. We hope that this investigation of efficient cooling and heat transferring techniques will contribute to further developments of quantum information technologies.

\begin{acknowledgements}
The numerical calculations were done by using the QuTiP library~\cite{Qutip1,Qutip2}. K.F. was supported by the JSPS KAKENHI Grant Number JP18J00454.  N.L. acknowledges partial support from JST PRESTO through Grant No. JPMJPR18GC. B.K. and J.P.P. acknowledge Academy of Finland grants 312057 and Marie Sklodowska-Curie actions (grant agreements 742559 and 766025). F.N. is supported in part by the: MURI Center for Dynamic Magneto-Optics via the Air Force Office of Scientific Research (AFOSR) (FA9550-14-1-0040), 
Army Research Office (ARO) (Grant No. W911NF-18-1-0358), Asian Office of Aerospace Research and Development (AOARD) (Grant No. FA2386-18-1-4045), 
Japan Science and Technology Agency (JST) (via the Q-LEAP program, and the CREST Grant No. JPMJCR1676), Japan Society for the Promotion of Science (JSPS) (JSPS-RFBR Grant No. 17-52-50023, and JSPS-FWO Grant No. VS.059.18N). F.N. and N.L. also acknowledge support from the RIKEN-AIST Challenge Research Fund, and the John Templeton Foundation.
\end{acknowledgements}

\end{document}